\documentclass[a4paper, pra, twocolumn, showpacs]{revtex4}

\usepackage{amssymb}
\usepackage{amsmath}
\usepackage{color}
\usepackage{graphics, graphicx}
\usepackage{bbold}

\begin{document}

\author{Michiel Snoek}
%\author{Masudul Haque}
\author{S. Vandoren}
\author{H. T. C. Stoof}
\affiliation{Institute for Theoretical Physics, Utrecht University, Leuvenlaan
4, 3584 CE Utrecht, The
Netherlands}
\date{\today}

\title{Theory of Ultracold Superstrings}

\begin{abstract}
The combination of a vortex line in a one-dimensional optical lattice with
fermions bound to the vortex core makes up an ultracold superstring. We give a
detailed derivation of
the way to make this supersymmetric string in the laboratory. In particular, we
discuss the presence of a fermionic bound state in the vortex core and the
tuning of the laser beams needed to achieve supersymmetry. Moreover, we discuss
experimental consequences of supersymmetry and identify the precise
supersymmetry in the problem. Finally,  we make the mathematical connection with string
theory.
\end{abstract} 

\maketitle

\section{Introduction}
Ultracold quantum gases provide a very exciting branch of physics. Besides
the interesting physics that the gases offer by themselves, it has also been
possible in
the last few years to model with quantum gases systems from other branches of
physics, and by doing so to
provide answers to long-standing questions. The latter is mainly due to the
amazing
accuracy by which their properties can be tuned and manipulated. This involves
the trapping potential, the dimensionality, the interaction between the atoms,
and the statistics. By using a
three-dimensional optical lattice the superfluid-Mott insulator transition in the
Bose-Hubbard model has been
observed \cite{Greiner02}. Bosonic atoms confined in one-dimensional tubes by
means of a two-dimensional optical lattice where shown to realize the
Lieb-Liniger gas \cite{Paredes04,Kinoshita04}. The unitarity regime of strong
interactions was reached by using Feshbach resonances to control the
scattering length \cite{Regal04, Zwierlein04, Kinast04, Bartenstein04, Bournel04}.

To this shortlist of examples
from condensed-matter theory, also examples from high-energy physics can be
added. In a spinor Bose-Einstein condensate with ferromagnetic interactions
skyrmion physics has been studied \cite{Khawaja01, Ruostekoski01},
whereas an antiferromagnetic
spinor Bose-Einstein condensate allows for monopole or hedgehog solutions \cite{Stoof01,
Martikainen02}.
There is also a proposal for studying charge fractionalization in one dimension
\cite{Ruostekoski02}, and 
for creating (static) non-abelian gauge fields \cite{Osterloh05, Ruseckas05}.
In recent work \cite{Snoek05} we have added another proposal to model a system
from high-energy physics. 
By combining a vortex line in a one-dimensional optical lattice with a
fermionic gas bound to the vortex core, it is possible to tune the laser
parameters such that a nonrelativistic supersymmetric string is created. This
we called the ultracold superstring. This proposal combines three topics that
have 
 attracted a lot of attention in the area of ultracold atomic gases.
These topics are vortices \cite{Matthews99, Madison00, Anderson00,
Hodby02, Fetter01_JP},
Bose-Fermi mixtures \cite{Hadzibabic02, Roati02, Inouye04,
Schori04, Goldwin04, Stan04, Silber05}, and optical lattices
\cite{Greiner02, Stoferle04}. 
Apart from its potential to experimentally probe certain aspects of superstring
theory, this proposal is also very interesting because it brings supersymmetry
within experimental reach.

Supersymmetry is a very special symmetry, that relates fermions and bosons with
each other. It plays an important role in string theory, where supersymmetry is
an essential ingredient to make a consistent theory without the so-called
tachyon, i.e., a
particle that has a negative mass squared.  In the physics of the minimally extended standard model,
supersymmetry is used to remove quadratic divergences. This results in a
super partner for each of the known particles of the standard model. However,
supersymmetry is manifestly broken in our world and none of these superpartners
have been observed. 
A third field where supersymmetry plays a role is in modeling disorder and
chaos \cite{Efetov}. Here supersymmetry is introduced artificially to properly
perform the average over disorder. 
Finally, supersymmetry plays an important role in the field of supersymmetric
quantum mechanics, where the formal structure of a supersymmetric theory is
applied to derive exact results. In particular this means that a supersymmetry
generator $Q$ is defined, such that the hamiltonian can be written as
$\mathcal{H} = \{Q, Q^\dagger\}$, which is one of the basic relations in the
relativistic superalgebra.
It is important for our purposes to note, that this relation is no longer enforced by the
superalgebra in the nonrelativistic limit. Careful analysis \cite{Puzalowski78, Clark84}
shows that in this limit the hamiltonian is replaced by the number operator,
i.e., $\mathcal{N} =  \{Q, Q^\dagger\}$. It may sometimes be possible to write a
nonrelativistic hamiltonian as the anticommutator of the supersymmetry
generators, but this does not correspond to the nonrelativistic limit of a
relativistic theory.

In our proposal, a physical effect of supersymmetry is that  the stability of the
superstring against spiraling out of the gas is
exceptionally large,
because the damping of the center-of-mass motion is reduced by a
destructive
interference between processes that create two additional bosonic excitations of
the superstring and processes that produce an additional particle-hole pair of
fermions.
Moreover, this system allows for the study of a quantum phase transition that
spontaneously breaks supersymmetry as we will show.

Another very interesting aspect of the ultracold superstring is the close
relation with string-bit models \cite{Bergman95}. These are models that
discretize the string in
the spatial direction, either to perturbatively solve string theory, or, more
radically, to reveal a more fundamental theory that underlies superstring
theory. String-bit models describe the transverse degrees of freedom of the
string in a very similar fashion as in our theory of the ultracold superstring.

In this article we investigate in detail the physics of ultracold
superstrings, expanding on our previous work \cite{Snoek05}.
The article is organized as follows. In Sec. II we give the detailed derivation
of the conditions for the ultracold superstring to be created. In particular, we
pay attention to the presence of the fermionic bound state in the vortex core
and the tuning of the lasers to reach supersymmetry. In Sec. III we investigate
the experimental consequences of the supersymmetry. Sec. IV contains a detailed
description of the supersymmetry by studying the superalgebra. In
Sec. V we make connection with string theory. Finally, we end with our conclusions
in Sec. VI. 

\section{Ultracold superstrings}
\begin{figure}
\hspace{-1cm}
\includegraphics[scale=.45]{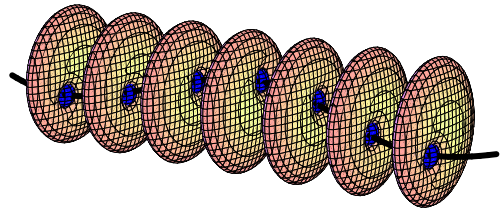}
\caption{(Color online) Artist's impression of the setup. The disks represent
the bosonic condensate density and the blue balls in the vortex core represent
the fermionic density. The black line is a guide to the eye to see the wiggling
of the vortex line that corresponds to a Kelvin mode.}
\label{artimpl}
\end{figure}

Our proposal makes use of the fact that a vortex line through a
Bose-Einstein condensate in a one-dimensional optical lattice can behave
according to the laws of quantum mechanics \cite{Martikainen03}. Such an
optical lattice consists of two identical counter-propagating laser beams and
provides a periodic potential for atoms. When applied along the symmetry axis of
a cigar-shaped condensate, which we call the $z$ axis from now on, the optical
lattice divides the condensate into weakly-coupled pancake-shaped condensates.
In the
case of a red-detuned lattice, the gaussian profile of the laser beam provides
also the desired trapping in the radial direction. Rotation of the Bose-Einstein
condensate along the $z$ axis creates a vortex line that passes through each
pancake. Quantum fluctuations of the vortex position are greatly enhanced in
this configuration because of the small number of atoms $N_B$ in each pancake,
which can be as low as $N_B=10$, but is typically around $N_B=1000$. 
An added advantage of the stacked-pancake configuration, as opposed to the bulk
situation, is that the dispersion of the vortex oscillations is particle like. 
This
ultimately allows for supersymmetry with the fermionic atoms in the mixture. In
the one-dimensional optical lattice the vortex line becomes a chain of so-called
pancake vortices. This produces a setup which is pictured schematically in Fig
\ref{artimpl}. 
There is a critical external rotation frequency $\Omega_c$  above which a vortex
in the center of the condensate is stable. For $\Omega < \Omega_c$ the vortex is
unstable, but because of its Euler dynamics, it takes a relatively long time
before it spirals out of the gas \cite{Anderson00, Fedichev99,
Duine04}. We analyze in detail the case of $\Omega=0$, i.e., the situation
in which the condensate is no longer rotated externally after a vortex is
created. However, the physics is very similar for all $\Omega < \Omega_c$, where
supersymmetry is possible.
The temperature is taken to be well below the Bose-Einstein condensation
temperature, so that thermal fluctuations are strongly suppressed. We only
consider the zero-temperature limit, 
because supersymmetry is formally broken for nonzero temperatures.

\subsection{Atomic species}
A convenient choice for the boson-fermion mixture is $^{87}$Rb and $^{40}$K,
since
such Bose-Fermi mixtures have recently been realized in the laboratory
\cite{Roati02, Inouye04, Schori04, Goldwin04}, and because the
resonance lines in these two atomic species lie very nearby. 
The mostly used $|f, m_f\rangle$ hyperfine spin states  are $| 9/2, \pm 9/2 \rangle$ and  
$|9/2, \pm 7/2\rangle$ for $^{40}$K, and
$|2, \pm 2 \rangle$, $|1, \pm 1 \rangle$, and $| 1,0 \rangle$ for $^{87}$Rb
\cite{Simoni03}. They all have a
negative interspecies scattering lenght $a_{BF}$, which is not desirable for our purposes as we show below. 
It could be possible to
use other spin states, which have a positive interspecies scattering
length. An other possiblity is to tune
the scattering length, using one of the various broad Feshbach resonances that
can make the interaction repulsive while keeping the probability to create
molecules negligible \cite{Simoni03}.

In principle it is also
possible to use other mixtures. Another Bose-Fermi mixture that has been
realized in the laboratory consists of $^{23}$Na and $^{6}$Li atoms
\cite{Hadzibabic02, Stan04}. This mixture is less convenient because the
resonance lines are widely separated, so that the two species feel very
different optical potentials and it is hard to trap both with a single laser. In
addition, $^6$Li is relatively hard to trap in an optical lattice because of its
small mass. For these reasons, the $^{23}$Na-$^6$Li mixture can only be used in
a very restricted parameter regime, as we will show lateron in Fig. \ref{tuning1}. 
For the same reasons, the
mixture $^{87}$Rb - $^6$Li \cite{Silber05} does not work well either. The
mixture $^7$Li-$^6$Li \cite{Truscott01, Schreck01} cannot be used at
all,
because the resonance lines of the species are the same, so it is impossible to
tune the physical properties of the mixture.

\subsection{Optical lattice}
\begin{figure}
\begin{center}
\begin{minipage}[b]{.6cm} (a) \vspace{2.9cm} \end{minipage} 
\hspace{-.9cm}
\includegraphics[scale=.65]{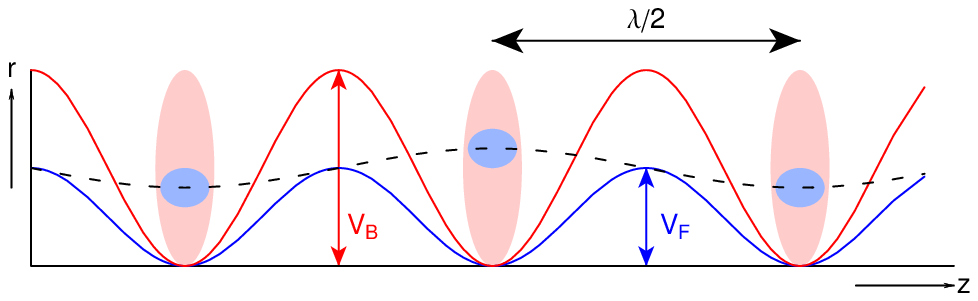}
\hspace{-.7cm}
\begin{minipage}[b]{.4cm} (b) \vspace{2.9cm} \end{minipage} 
\hspace{-.32cm}
\includegraphics[scale=.4]{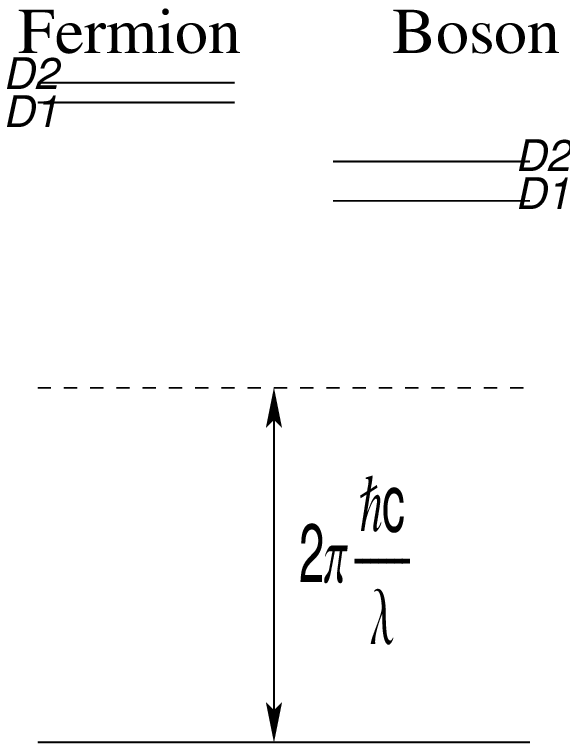}
\end{center}
\caption{(a) (Color online) Schematic picture of the setup. Here $r$ is the
radial distance in the
$xy$ plane. The pink and blue blobs represent the bosonic and fermionic
densities, respectively. Moreover, $\lambda$ is the wavelength of the laser. The
blue and red lines indicate the strength of the optical potential, respectively,
for the bosons and fermions as a function of the $z$ coordinate. (b) Schematic
fine structure level scheme of the bosonic and fermionic atomic species. Because
we consider only sufficiently large detunings the hyperfine level structure is
not resolved.}
\label{optpotential}
\end{figure}

Because the excited states of the bosonic and fermionic atoms have different
transition frequencies, 
the optical lattice produces for the two species a periodic potential with the
same lattice spacing,
but with a different height, as schematically shown in Fig. \ref{optpotential}. 
This is very crucial, because it allows to tune the optical lattice for the
bosonic and fermionic atoms seperately, by careful adjustement of the wavelength
and the Rabi
frequency, i.e., the intensity of the laser. This is required
to be able to tune the system to become supersymmetric lateron.

For the $^{86}$Rb-$^{40}$K
mixture the Rabi frequencies are in a good approximation the same. For other
mixtures
the Rabi frequencies are different and we then take the bosonic Rabi frequency
as a reference.
We take into account the fine-structure level scheme of the atoms, but, assuming
that we are sufficiently far from resonance,
we neglect the hyperfine structure. As a result, the optical potential is given
by
\begin{equation}
		V_{B,F} (z) =  V_{B,F} \cos^2 (2\pi z/\lambda),
\end{equation}
where the well depths obey
\begin{eqnarray}
	V_{B,F}&=&- \frac{\hbar \Omega_{B,F}^2}{3} \left[
		 \left(  \frac{1}{\omega^{B,F}_{D_1} - \omega }
		+ \frac{1}{\omega^{B,F}_{D_1} + \omega} \right) \right. \\
	&&	\hspace{1cm} \left.+ 2
		\left( \frac{1}{\omega^{B,F}_{D_2} - \omega} +
		\frac{1}{\omega^{B,F}_{D_2} + \omega}
		\right) \right],
	\nonumber
\end{eqnarray}
$\omega= 2 \pi c/\lambda$ is the laser frequency, and $\omega_{D_1}$ and
$\omega_{D_2}$ are the frequencies of the $D_1$ and $D_2$ resonance lines. Here
we neglected spontaneous emission of photons. This effect slightly modifies the
trapping potential, but gives a finite lifetime to the atoms. Using the
rotating-wave approximation and neglecting the fine structure, the effective rate of
photon
absorption can for red-detuned laser light be estimed as
\begin{equation}
	\Gamma_{B,F}^{\rm eff}= - \frac{\hbar \Omega_{B,F}^2}{2} 
		  \frac{\Gamma_{B,F} }{(\hbar \omega^{B,F} - \hbar \omega)^2 +
(\hbar \Gamma_{B, F})^2   },
\end{equation}
where $\Gamma_{B,F}$ is the linewidth of the bosonic or fermionic excited state,
respectively. For blue-detuned
laser light, the atoms are trapped in the regions of low laser intensity and
spontaneous emission is strongly reduced.

The optical potential should be sufficiently deep to have a bound state for the bosonic and fermionic
atoms. To make sure that that is the case
we impose the condition
\begin{equation}
		\frac{V_{B,F}}{E_{B,F}} > \frac{3}{2},
\end{equation}
where we have used the recoil-energy 
\begin{equation}
		E_{B,F} = \frac{2 \pi^2}{m_{B,F} \lambda^2},
\end{equation}
which is the energy associated with the absorption of a photon.
On the other hand, the optical lattice should not be so strong to drive the
system in the Mott-insulator state \cite{Greiner02, Stoferle04, Oosten03}. In
one dimension with many atoms per site, this requires an exceptionally deep
lattice, which only occurs
if the laser frequency is very close to the resonance frequency of the atomic
species. Since we stay away from resonance, this
situation does not occur in our calculations.

The wavefunctions in the $z$ direction are assumed to be the groundstate wavefunctions of the
harmonic oscillator associated with the optical lattice and thus given by
\begin{equation}
\psi_{B,F} (z) = \frac{1}{\pi^{1/4} \sqrt{\ell_{B,F}^z}} \exp
\left(-\frac{z^2}{2
\ell_{B,F}^z} \right),
\end{equation}
where
\begin{equation}
\ell_{B,F}^z = \left(\frac{E_{B,F}}{V_{B,F}}\right)^{1/4} \frac{\lambda}{2 \pi}.
\end{equation}
For the tunneling amplitude, we use the expression \cite{abromowitz}
\begin{equation}
		J_{B,F} = 4 \frac{(V_{B,F}^{3} E_{B,F})^{1/4}}{\sqrt{\pi}}
\exp\left[-2
		\sqrt{V_{B,F}/{E_{B,F}}}\right],
\end{equation}
which becomes exact for a deep lattice. Therefore, the atomic dispersions along the $z$ axis are
given by
\begin{equation}
		\epsilon_{B,F}(k)=2 J_{B,F}\lbrack 1-\cos(k \lambda/2) \rbrack.
\end{equation}
Lateron we need for the fermions the relation between the average number
of particle per site and
the chemical potential $\mu_F$. From the above dispersion we derive at zero
temperature that
\begin{equation}
N_F = \frac{2}{\pi} \arcsin \left[ \sqrt{\frac{\mu_F}{4 J_f}}\right],
\label{fillingfrac}
\end{equation}
where we neglect also interaction effects.

%\subsection{Rotation}
%\subsection{Temperature}
%The temperature is taken to be well below the Bose-Einstein condensation
%temperature, so that thermal fluctuations are strongly suppressed. 

\subsection{Kelvons}
The wavefunction in the (axial) $z$ direction is fully specified by the optical
lattice and all the dynamics thus takes place in the radial direction, i.e., in
the 
$xy$-plane. Since the vortex-fluctuations form the lowest-lying modes, we
restrict
the dynamics to the vortex motion.
We follow the derivations in earlier work \cite{Martikainen03,
Martikainen04}, 
where a specific ansatz for the wavefunction was used, to achieve this. In this
work the condensate density was described by a gaussian wavefunction with size
$R_{\rm TF}$ and the vortex core was approximated by a step function. Furthermore, it was
assumed that the vortex is close to the center.
The motion of the vortices results in 
kelvons, i.e., quantized
oscillations of the vortex, described by the creation and annihilation
operators
\begin{equation}
\hat b=\frac{\sqrt{N_b}}{R_{\rm TF}}(\hat x +  i \hat y), \quad \hat b^\dagger =
\frac{\sqrt{N_b}}{R_{\rm TF}}(\hat x -  i \hat y) ,
\end{equation}
which obey $
\lbrack \hat b, \hat b^\dagger \rbrack =1$.
Without the optical lattice Kelvin waves have already been observed \cite{Bretin03, Mizushima03}.
The kelvons have the dispersion
\begin{eqnarray}
		\hbar \omega_K(k) &=& \frac{\hbar \omega  \ell^2}{2
R_{\rm TF}^2}\left(1-\Gamma\left[0, \left(\frac{l}{R_{\rm TF}}\right)^4 \right] \right) + \hbar
\Omega  
\label{dispersion} \\ &&+
2 J_K \lbrack 1-\cos(k \lambda/2)\rbrack, \nonumber 
\end{eqnarray}
where
\begin{equation} \label{jk}
		J_K = \Gamma[0, (\ell/R_{\rm TF})^4]J_B,
\end{equation}
$\Gamma[0,z]$ is the incomplete Gamma function, $R_{\rm TF}$ is the Thomas-Fermi
radius in the radial direction,
$\ell$ is the bosonic harmonic length in the radial direction, and $\omega$  is
the
associated frequency.
Using another ansatz for the condensate wavefunction can slightly change the constant of
proportionality in the definition of the kelvon operators and in the details of
the dispersion, but the dispersion always stays tight-binding like. 
%These change
%are thus mainly quantitative, and do not bring in new physics. 
%However, these quantitative differences can 
%be important in the actual realization of the experiment. For this moment we
%will use these results, whereas later in the %paper we will investigate whether
%another ansatz for the wavefunction makes an improvement.

For the calculation of the bound state in the vortex core, we need to go beyond
the description of the core by a step function. This change
of the calculation could improve the value of $J_K$, but not the functional form
of the kelvon dispersion. Since the corrections on the value of $J_K$ are
small, we just use
the result in Eq. (\ref{jk}). Besides the bandwidth $J_K$ we derive from
Eq.\ (\ref{dispersion}) also
the chemical potential for the kelvons, which gives
\begin{eqnarray}
\mu_K &\equiv& - \hbar \omega_K(k=0) \\
&=& \frac{\hbar \omega  \ell^2}{2
R_{\rm TF}^2}\left(\Gamma\left[0, \left(\frac{l}{R_{\rm TF}}\right)^4 \right] -1 \right) - \hbar
\Omega . \nonumber
\end{eqnarray}
Note that the chemical potential is positive only for sufficiently small rotations, which is due to the fact that the vortex is in principle unstable for these values of the rotation and wants to spiral out of the center of the gas cloud.

\subsection{Bound states in the vortex core}
By treating the interaction between the bosonic and fermionic atoms in
mean-field
approximation, we have to solve the Gross-Pitaevskii equation for the condensate
wavefunction $\Psi({\bf r})$ coupled to the Schr\"odinger equation for the
fermion
wavefunction $\psi({\bf r})$
\begin{widetext}
\begin{eqnarray}
	(-\frac{ \hbar^2 \nabla_{\bf r}^2}{2 m_B} + \frac{1}{2} m_B \omega_B^2
r^2 -
\mu_B + \frac{U_{BB}}{2} | \Psi({\bf r}) |^2
	+ U_{BF} |\psi({\bf r})|^2 - \hbar \Omega \hat L_z) \Psi({\bf r}) &=& 0, \\
	(-\frac{ \hbar^2 \nabla_{\bf r}^2}{2 m_F} - E + \frac{1}{2} m_F
\omega_F^2 r^2
	+ U_{BF} |\Psi({\bf r})|^2) \psi({\bf r}) &=& 0,
\end{eqnarray}
\end{widetext}
which we investigate for the case that $\Omega=0$.
The interaction paramaters are related to the scattering lengths according to
\begin{eqnarray}
U_{BB} &=&
\frac{4 \pi \hbar^2}{\sqrt {2 \pi} m_B} \frac{a_{BB}}{\ell^z_B}, \\
U_{BF} &=& \frac{2 \pi \hbar^2}{\sqrt {\pi} m_R}
\frac{a_{BF}}{\sqrt{(\ell^z_B)^2+(\ell^z_F)^2}},
\end{eqnarray}
with $a_{BB}$ the boson-boson scattering length and $a_{BF}$ the boson-fermion
scattering length and
$m_R$ the reduced mass $m_B m_F/(m_B+m_F)$.
Although it is very well possible to solve these equations numerically, we
prefer an analytic treatment, to gain more insight into the problem. To proceed
we
make the approximation that the condensate wavefunction is not affected
by the presence of the fermions. This is justified, because the contribution of
the
fermions is ${N_F}/{N_B}$ smaller
than the contribution of the bosons, where $N_{B,F}$ is the average number of
bosons and fermions at a lattice site.
This ratio will be smaller than $10^{-3}$ as it turns out. Taking into account
the interaction with the fermions leads to a slightly
wider vortex core, which enhances the possiblity of a bound state.
So we first solve the Gross-Pitaevskii equation for the condensate
density neglecting the
presence of the fermions and then use the condensate density as an effective
potential for the fermions. Since we only
want to estimate when there is a bound state and we do not need the details of
this bound state, we make the following
approximations.
First, we assume the vortex to be in the center such that the problem is
rotationally symmetric and we only have to solve the radial equation. Because of
the quantum uncertainty the vortex position in principle fluctuates around the
center of the trap,
but these fluctuations are small.
Second, we assume the envelope condensate wavefunction to be Thomas-Fermi like,
i.e.,
\begin{equation}
|\Psi({\bf r})|^2=
n_{TF} ({\bf r}) = \frac{\mu_B - \frac{m_B}{2} \omega_B^2 r^2}{U_{BB}}
 = n_0 \left(1-\frac{r^2}{R_{\rm TF}^2}\right).
\end{equation}
Third, we describe the vortex core by ${r^2}/(r^2+ 2 \xi^2)$
\cite{Baym04}, such that the total bosonic
density is given by
\begin{equation}
n_B ({\bf r}) = \left(\frac{r^2}{r^2+ 2 \xi^2} \right) n_{TF} ({\bf r}).
\label{vortexcore}
\end{equation}
If we take for the healing length $\xi$ the usual expression in the center of
the trap, i.e.,
\begin{equation}
\xi = \frac{\hbar}{\sqrt{2 m_B n_{TF} (0) U_{BB}}},
\end{equation}
we obtain the relation
\begin{equation}
\frac{\xi}{\ell} = \frac{\ell}{R_{\rm TF}}.
\end{equation}
By expressing the energy in terms of $\mu_B m_B/m_F$ we can write the
Schr\"odinger
equation
for the fermions as
\begin{widetext}
\begin{equation}
	\left[-\frac{ \xi^2 \nabla_r^2}{2} - \epsilon +  \frac{r^2
\xi^2}{(\ell_F^r)^4}
	+ \frac{m_F}{m_B} \frac{U_{BF}}{U_{BB}} \frac{r^2}{2
\xi^2+r^2}\left(1-\frac{r^2 \xi^2}{\ell^4}\right)\right] \psi(r) = 0,
\end{equation}
\end{widetext}
where $\epsilon = m_F E/\mu_B m_B$ and the dimensionless parameter
\begin{equation}
\Gamma=\frac{m_F}{m_B}
\frac{U_{BF}}{U_{BB}}=\frac{1}{2}\sqrt{\frac{2}{1+\sqrt{\frac{V_B}{V_F}
\frac{E_F}{E_B}}}} \left(1+\frac{m_F}{m_B}\right) \frac{a_{BF}}{a_{BB}}
\end{equation}
determines whether or not there is a bound state in the core of the vortex.

If we assume that $\xi \ll \ell$ and $\xi \ll \ell_r^F$, we can neglect the harmonic 
confinement and the Thomas-Fermi profile of the Bose-Einstein condensate. The effective potential for the fermionic atoms is then given by $\Gamma r^2/(r^2 + 2 \xi^2)$. This potential has a bound state for each value of $\Gamma$, because for large distances from the core, it behaves as $\Gamma(1-2 \xi^2/r^2)$. However, the size of the wavefunction describing the bound state becomes extremely large for small values of $\Gamma$. 
Hence it is necessary to take into account the exact form of the potential to make a quantitative estimate of the existence of the bound state. The potential is determined by the values of the radial bosonic and fermionic harmonic length $\ell$ and $\ell_F^r$. Since $\ell_F^r$ determines the potential outside the condensate, it determines whether or not the fermions can tunnel out of the core to this region. For the existence of the bound state we can neglect this contribution, which is always justified, because it enhances the possibility of having a bound state.
 
The radial bosonic harmonic length $\ell$ is fixed by the normalization of the
condenstate wavefunction
\begin{equation}
\int d^2{\bf r} |\Psi({\bf r})|^2 = \int d^2{\bf r} \frac{r^2}{r^2+ 2 \xi^2}
n_{TF} ({\bf r})= N_B.
\end{equation}
Neglecting the presence of the core we find the usual expression for the 
Thomas-Fermi profile
\begin{equation}
\frac{\ell^4}{\xi^4}  = \frac{R_{\rm TF}^4}{\ell^4} =\frac{4 m_B N_B U_{BB}}{\pi \hbar^2}=
 \frac{16 \sqrt{2 \pi} N_B a_{BB}}{\lambda} 
\left[\frac{V_B}{E_B}\right]^{1/4} \! .
\label{TFR}
\end{equation}
Using that 
\begin{eqnarray*}
&& \int d^2 {\bf r} \left(\frac{r^2}{r^2 + 2 \xi^2}\right)\left( 1 -
\frac{r^2}{R_{\rm TF}^2} \right) = \\ 
&& \hspace{.1cm} \frac{\pi}{2} R_{\rm TF}^2 + 2 \pi \xi^2 \left(1 - \left[1+ \frac{2
\xi^2}{R_{\rm TF}^2}\right]\log \left(\frac{R_{\rm TF}^2 + 2 \xi^2}{2 \xi^2} \right)\right), 
\end{eqnarray*}
we see that taking into account the core implies that we have to solve the equation
\begin{equation}
\frac{\ell^4}{\xi^4}= \frac{4 m_B N_B U_{BB}}{\pi \hbar^2} + 4 \left(1+\frac{2
\xi^4}{\ell^4}\right)\log\left(1+\frac{\ell^4}{2 \xi^4}\right) - 4,
\end{equation}
where the last two terms come form the presence of the core.
Since the core is small in this approximation, this results in a radial harmonic
length that is only slightly modified. 
The requirement that the wavefunction should vanish well within the condensate can then be quantified 
to yield the expression 
\begin{equation} 
\Gamma \frac{\sigma^2}{\sigma^2 + 2 \xi^2} \left(1 - \frac{\sigma^2 \xi^2}{ \ell^4  }\right) < \epsilon,
\end{equation}
where $\sigma$ is the radial size of the fermionic wavefunction.
In this way we obtain that for typical densities there is a bound state for $\Gamma>1.5$ which means that
${a_{BB}}/{a_{BF}}>2$.

In contrast to the radial bosonic length $\ell$, the fermionic radial harmonic
length $\ell_F^r$ is not fixed. 
When the optical lattice is red-detuned, the lattice can be used to trap the atoms also
in the radial direction. As a consequence, the total confining potential for the
fermions is a multiple of the confining potential of the bosons, i.e.,
\[
\frac{r^2}{\ell^2} + \frac{z^2}{(\ell_B^z)^2} \propto \frac{r^2}{(\ell_F^r)^2}
+ \frac{z^2}{(\ell_F^z)^2}.
\]
This gives the relation
\begin{equation}
\frac{\ell_F^r}{\ell}=\frac{\ell_F^z}{\ell_B^z},
\end{equation}
from which we derive
\begin{equation}
\left(\frac{\ell_F^r}{\xi}\right)^4=\left(\frac{\ell_F^z}{\ell_B^z}\frac{\ell}
{\xi}\right)^4 =
\left(\frac{\ell}{\xi}\right)^4 \frac{E_F}{E_B}\frac{V_B}{V_F}.
\end{equation}

\begin{figure}
\hspace{3cm}
\includegraphics[scale=.25]{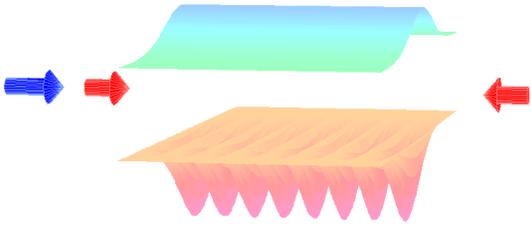}
\caption{(Color online) Setup with the additional laser to manipulate the radial trapping
potential.}
\label{extra_laser} 
\end{figure}

\begin{figure}
\begin{center}
\includegraphics{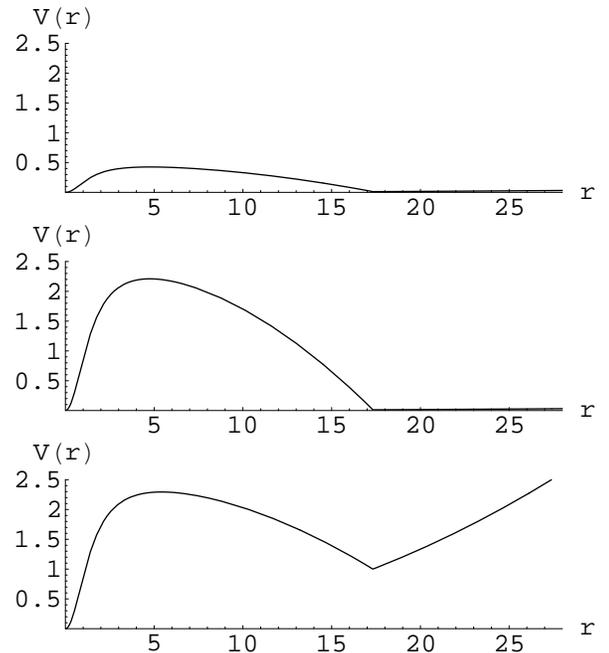}
\end{center}
\caption{Effective potential for the fermions. Lengths are measured in units of
$\xi$ and energies in units of $\mu_B m_B/m_F$. (a) $\Gamma=.5$, $\ell_F^r \gg
\ell$: No bound state, since the potential is too small.
 (b) $\Gamma=2.6$, $\ell_F^r \gg \ell$: Bound state in the core, but possibility
to tunnel outside.
  (c) $\Gamma=2.6$, $\ell_F^r \simeq \ell$: Bound state in the core, no tunneling possible.}
\label{effpot}
\end{figure}

However, if the lattice is blue-detuned or if the radial trapping is tuned
independently, this relation is not true. 
The radial trapping can be tuned by introducing a second running laser 
in the same direction as the optical lattice,
as shown in Fig \ref{extra_laser}.
The new laser beam has a constant intensity along the $z$ axis, and does not 
influence the one-dimensional
potential wells, but it does change the radial confinement. In principle this
second laser also
introduces interference terms, but they are much faster than the atoms can
follow for the frequencies of interest to us. Therefore, the intensities of the
two lasers can simply be added. 
In particular, as we show lateron, adjusting the radial trapping
potentials is needed to get supersymmetric interaction terms. The condition
imposed by this requirement is
\begin{equation}
	\left(\frac{\ell}{\ell_F^r}\right)^4=\frac{m_B}{m_F}
\left(\frac{\xi}{\ell}\right)^4
	\left(\Gamma\left[0,\left(\frac{\xi}{\ell}\right)^4\right]-
	\frac{3}{2}\right),
\end{equation}
which gives the following expression for the fermionic radial harmonic length
\begin{equation}
\left(\frac{\ell_F^r}{\xi}\right)^4 = 
\frac{m_F}{m_B}
\left(\frac{\ell}{\xi}\right)^8
\frac{1}
{\Gamma\left[0,\left(\frac{\xi}{\ell}\right)^4\right] - \frac{3}{2}}.
\end{equation}
In this last case, the harmonic radial potential for the fermions is very small.
In principle this allows the fermionic atoms to tunnel out of the vortex core,
to the region where the
condensate density vanishes. However, the tunneling is suppressed by increasing
the parameter $\Gamma$.
A WKB estimate gives that for $\Gamma>5$ the lifetime of the fermions in the core
is larger than a second.
This means that ${a_{BB}}/{a_{BF}}>8$. Further increasing this ratio
increases this lifetime dramatically. 
Since adjusting the radial trapping potentials is only needed close to
the center of the trap, it is also a
possibility to use a second laser with a much smaller waist, such that
higher-order contributions from the potential prevent the fermions from tunneling
out of the core. For
various situations, the effective potential for the fermions 
%and the fermionic wave function 
is shown in Fig
\ref{effpot}.

\subsection{Interactions}
In our superstring realization there are also boson-boson and boson-fermion
interactions. The kelvons interact repulsively among each other when $\Omega < 
\Omega_c$. For $\Omega=0$ the kelvon-kelvon interaction coefficient
is given by \cite{Martikainen04B}
\begin{equation}
V_{KK}= \frac{\hbar^2}{2 N_B m_B R_{\rm TF}^2}\left(\Gamma [0, \left({\ell}/{R_{\rm TF}}\right)^4
]-
\frac{3}{2}\right).
\end{equation}
In addition, a repulsive interaction between the kelvons and the fermionic atoms
is generated by the fact that physically the presence of a kelvon means that the
vortex core is shifted off center, together with the fermions trapped in it.
Because of the radial confinement experienced by the trapped fermions, this
increases the energy of the vortex. When the vortex core is
shifted from $0$ to $r$, the fermion hamiltonian is extended by a term
\begin{eqnarray}
\mathcal{H}_{KF} = \frac{ m_F (\omega_F^r )^2}{2} r^2 c^\dagger c,
\end{eqnarray}
where $c^\dagger c$ is the number operator for the fermions in the core.
Defining $C_{B,F}$ as the spring constants associated with the radial
confinement of the bosonic and fermionic atoms, respectively, and using the
definition of the kelvon operators, this translates into
\begin{eqnarray}
\mathcal{H}_{KF} = \frac{ C_F R_{\rm TF}^2}{2 N_B}  c^\dagger c \; b^\dagger b.
\end{eqnarray}
So the kelvon-fermion interaction coefficient is found to be 
\begin{equation}
V_{KF}= \frac{C_F R_{\rm TF}^2}{2 N_B}.
\end{equation}

\subsection{Supersymmetry}
\begin{figure}
\begin{center}
\includegraphics[scale=.65]{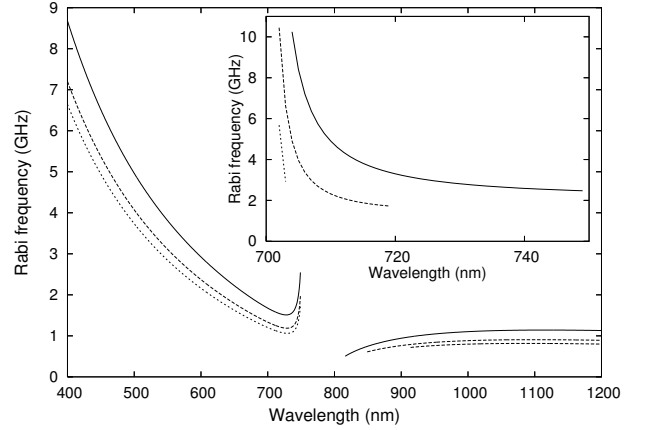}
\end{center}
\caption{Tuning of the lattice laser to obtain supersymmetry. Plotted is the
 Rabi frequency for the bosonic atoms versus wavelength
for $^{87}$Rb-$^{40}$K  
for 10000 (solid line), 1000 (dashed line) and 500 (dotted line) bosonic atoms per site.
Note that
for the blue-detuned part, i.e., $\lambda < 760$ nm for the  $^{87}$Rb-$^{40}$K
mixture extra radial trapping is needed, either magnetically, or by using an
extra running laser as discussed in the text and shown in Fig. \ref{extra_laser}. In Fig.
\ref{tuning3} we display how to tune the running laser to obtain also supersymmetric
interactions. In the inset we plotted  the Rabi frequency that is required for the $^{23}$Na-$^6$Li mixture to obtain supersymmetry, again for 10000 (solid line), 1000 (dashed line) and 500 (dotted line) bosonic atoms per site. Note that this can only be obtained in a very limited range of wavelength's.}
\label{tuning1}
\end{figure}

\begin{figure}
\begin{center}
\includegraphics[scale=.65]{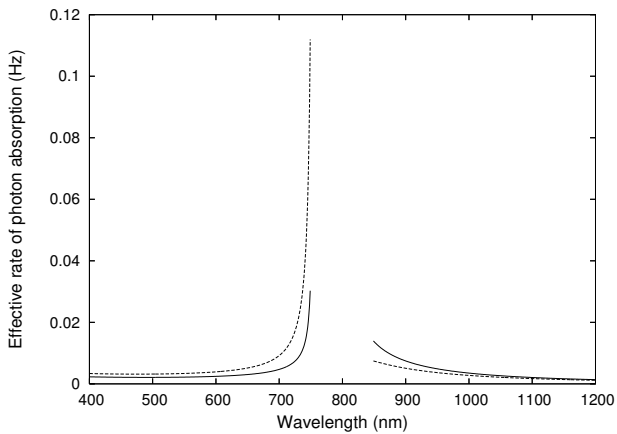}
\end{center}
\caption{Effective rate of photon absorption as a function of the wavelength of
the optical lattice for 1000 bosonic atoms per site. The solid line is for the
bosonic atoms, whereas the dashed line is for the fermionic atoms in the $^{87}$Rb-$^{40}$K mixture.}
\label{atomloss1}
\end{figure}

To obtain a supersymmetric situation we have three requirements. In the first
place the hopping
amplitudes have to be the same
\begin{equation}
J_F=J_K \equiv t.
\end{equation}
This can be done by adjusting the laser parameters $\lambda$ and
$\Omega_{B}$, as shown in Fig \ref{tuning1}. The freedom in choosing the
wavelength of the laser can be used to minimize the atom loss. In Fig.
\ref{atomloss1}, we plot the atom loss as a function of the wavelength of the
laser. 

\begin{figure}
\begin{center}
\includegraphics[scale=.65]{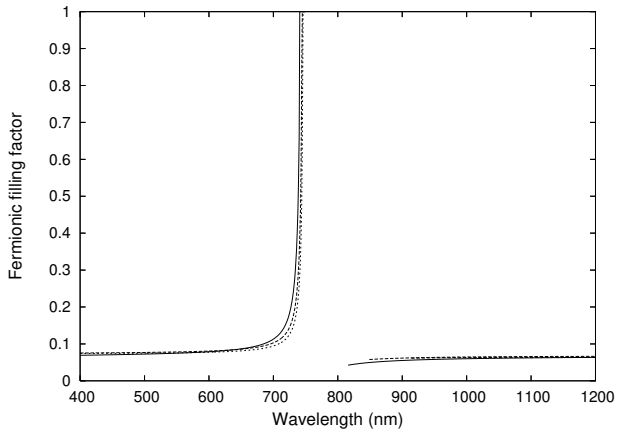}
\end{center}
\caption{Tuning of the average number of fermions per lattice site to obtain
supersymmetry for 10000 (solid line), 1000 (dashed line) and 500 (dotted line) bosonic atoms per lattice site. The result depends on the
ratio of the bosonic
harmonic lengths in the axial and radial directions $\ell/\ell_B^z$. This ratio
should be sufficiently small to be radially in the Thomas-Fermi limit. For this
plot a ratio of 1/5 is chosen.}
\label{tuning2}
\end{figure}

Secondly, the chemical potentials have to be the same
\begin{equation}
\mu_F= \mu_K \equiv \mu.
\end{equation}
This can be achieved by adjusting the fermion filling fraction $N_F$, as shown in
Fig \ref{tuning2}. Using the result from Eq. (\ref{fillingfrac}) and using the
requirements for
supersymmetry we obtain 
\begin{eqnarray}
&& N_F =   \frac{2}{\pi} \arcsin \left( \sqrt{ \frac{
\frac{\hbar \omega  \ell^2}{2
R_{\rm TF}^2}\left(\Gamma\left[0, \frac{l^4}{R_{\rm TF}^4} \right] -1 \right) }{4 J_B
\Gamma\left[0, \frac{l^4}{R_{\rm TF}^4} \right]}} \right) \nonumber \\
&& = \!  \frac{2}{\pi} \arcsin \! \left( \! \frac{\ell}{\ell_B^z} \sqrt{ \frac{
\sqrt{\pi} \frac{\ell^2}{
R_{\rm TF}^2} \left(\Gamma\left[0, \frac{l^4}{R_{\rm TF}^4} \right]\!- \! 1\! \right)
e^{\sqrt{\tfrac{V_B}{E_B}}}}{16 (V_B/E_B)^{1/4} \Gamma\left[0,
\frac{l^4}{R_{\rm TF}^4} \right]}} \right) \! . \nonumber
\end{eqnarray}
The ratio $\ell/\ell_B^z$ is undetermined by supersymmetry constraints. In
order for the Thomas-Fermi approximation to apply in the radial direction,
versus the gaussian wavefunction in $z$ direction, this ratio needs to be
sufficiently small. In the figure a ratio of $1/5$ is chosen.

\begin{figure}
\begin{center}
 \includegraphics[scale=.65]{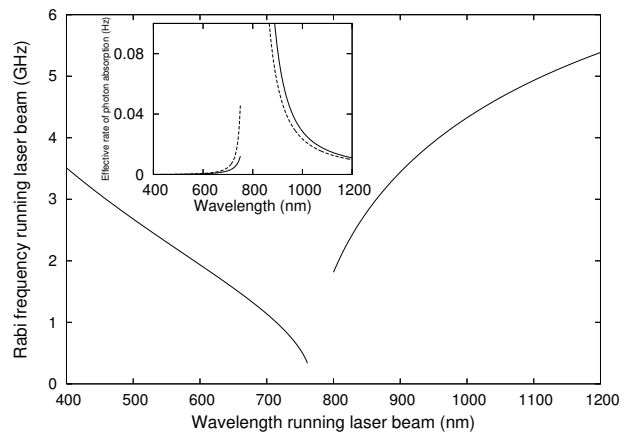}
\end{center}
\caption{Tuning of the additional laser to obtain supersymmetric interactions
for 1000 bosonic atoms per lattice site. The Rabi frequency is plotted versus the
wavelength of the running laser for different wavelengths of the lattice laser
beam: 1000 nm for the left curve and 600 nm for the right curve. In the inset the effective rate of photon absorption originating from the additional laser is plotted, again the left curves are for a lattice laser of 1000 nm and the right curves are for a lattice laser of 1000 nm. The solid lines indicate the rate of photon absorption for the bosonic atoms, whereas the dashed lines are the rates of photon absorption for the fermionic atoms.}
\label{tuning3}
\end{figure}

Finally, the interaction terms have to be the same. This implies
\begin{equation}
V_{KK}= V_{KF} \equiv U.
\end{equation}
Setting these coefficients equal to each other gives a condition on the radial
trapping given by
\begin{equation}
\frac{C_F}{C_B} =\left(\frac{\ell}{R_{\rm TF}} \right)^4 \left(\Gamma [0,
\left({\ell}/{R_{\rm TF}}\right)^4 ]-
\frac{3}{2}\right).
\end{equation}
The radial trapping can be tuned by introducing a second running laser, as
explained before. For the second laser, we can again independently choose both
the
wavelength and the Rabi frequency as shown in Fig. \ref{tuning3}.  This can again be used to
minimize the atom loss due to the red-detuned laser, but it turns out that atom
loss is always quite small anyway for reasonable system parameters. Only for very small detunings, the lifetime is less than a second.

\subsection{Hamiltonian}
Combining everything, our superstring is described by the supersymmetric hamiltonian
\begin{eqnarray} \label{hamiltonianl}
\mathcal{H} &=& - t \sum_{\langle i j \rangle} \lbrack b_i^\dagger b_j +
c_i^\dagger c_j \rbrack
  \\
&&
\nonumber \\ && + \sum_i  \Bigl\lbrack - \mu' ( b_i^\dagger b_i + c_i^\dagger
c_i) +
\frac{U}{2} b_i^\dagger b_i^\dagger b_i b_i+ U b_i^\dagger b_i 
c_i^\dagger c_i \Bigr\rbrack. \nonumber
\end{eqnarray}
Here $b_i$ is the annihilation operator of a kelvon at site $i$, $c_i$ is the
annihilation operator of a fermion at site $i$,  $\langle i j \rangle$  means
that the summation runs over neighbouring sites, and $\mu'= \mu - 2 t$. We used the convention for the
Fourier transformation
$f_k =
(1/\sqrt{N_s})
\sum_n e^{i k z_n} f_n$, where $N_s$ is the number of lattice sites.
We define $a=\lambda/2$ as the lattice spacing and $L=N_s a$ as the length of
the system. Assuming that $N_s \gg 1$, such that
$L \gg a$, we can 
perform a continuum approximation to obtain for the hamiltonian 
\begin{eqnarray}
\mathcal{H} &=& \int d z  \; 
b^\dagger (z) \left( - \frac{\hbar^2}{ 2 m^*} \frac{\partial^2}{\partial z^2} -
\mu \right)b(z) 
\label{ham_con}
\\ && 
+ \int d z \;  c^\dagger (z) \left( - \frac{\hbar^2}{2
m^*}\frac{\partial^2}{\partial z^2} - \mu
\right)c(z) 
\nonumber \\ && + \frac{U}{2} \int d z \; \lbrack
 b^\dagger (z) b^\dagger (z) b(z) b(z) + 2 b^\dagger(z) b(z)
c^\dagger (z) c(z) \rbrack, \nonumber 
\end{eqnarray} 
where we introduced the effective mass $m^* = \hbar^2/2 a^2 t$. 
This continuum hamiltonian 
turns out to be exactly solvable \cite{Batchelor05,  Imambekov06} by a
straightforward generalization of 
the Bethe-Ansatz solution of the one-dimensional Bose gas \cite{Lieb63,
Yang67}.
However, the exact solutions spontaneously break supersymmetry and do not give
much insight in the
role of supersymmetry in the problem.

Using that the lagrangian is given by
\begin{equation} \label{lagrangian}
\mathcal{L} = \sum_i \left( b_i^\dagger i \hbar \frac{\partial}{\partial t} 
b_i + 
c_i^\dagger i \hbar \frac{\partial}{\partial t} c_i \right)  
- \mathcal{H}[b^\dagger, b; c^\dagger, c],
\end{equation}
the action in the continuum limit is obtained as
\begin{eqnarray}
S \!\! &=& \!\! \int \! \! d t \! \int \!\! d z \left\{ b^*  \! \left( \! i
\hbar \frac{\partial}{\partial t} + \frac{\hbar^2}{2 m^*} \frac{\partial^2
}{\partial z^2} + \mu  \! \right) \! b \right. \\
&& \hspace{1.1cm}
+ c^* \left( \! i \hbar \frac{\partial}{\partial t} + \frac{\hbar^2}{2 m^*}
\frac{\partial^2 }{\partial z^2} + \mu  \! \right) \! c \nonumber \\
&& \hspace{1.5cm}
- \left. \frac{U}{2}\left(|b|^2 + |c|^2 \right)^2 \right\},
\nonumber
\end{eqnarray}
which now explicitly shows the supersymmetry of the problem, because it remains invariant
when $b$ and $c$ are rotated into each other.
If we neglect the interaction terms, which are rather small anyway,
the fermions fill a Fermi sea and the low-energy excitations are particle-hole
excitations around the Fermi surface.
Therefore, the low-energy part of the theory is properly described by
linearizing the fermionic dispersion around the Fermi level. To preserve
supersymmetry we do the same for the bosons and obtain at the quadratic level
the action
\begin{eqnarray}
S \!\! &=& \!\!  \sum_{\sigma = \pm } \! \int \! dt \!\! \int \! d k  \Bigl\{ b_\sigma^*
(k, t) ( i \hbar
\partial_t - \hbar v_F ( \sigma k - k_F) ) b_\sigma (k, t) \nonumber \\ && \hspace{1.7cm} + 
c_\sigma^* (k, t)( i \hbar \partial_t - \hbar v_F ( \sigma k - k_F)) c_\sigma (k, t)
\Bigr\},
\nonumber
\end{eqnarray}
where $\sigma$ indicates whether the particles are right movers or left movers
and $v_F = \hbar k_F/m^*$ is the Fermi velocity. We used that $\mu= \hbar^2 k_F^2/2
m^*$. We identify the 
Fermi velocity with the velocity of light $c$ and perform the transformation
$c_\sigma(z, t) = (1/\sqrt{L}) \sum_k e^{i k z} c_\sigma(k+ \sigma k_F, t)$.
We introduce the Dirac spinor $\psi(z,t) = ( c_+(z,t), c_-(z,t))$ and $\bar \psi
(z,t) = \psi^\dagger (z,t) \gamma^0$, with $\gamma^0 = \sigma_y$. The other
Dirac matrices are $\gamma^1= i \sigma_x$ and $\gamma^5 = \sigma_z$. 
The two bosonic fields can be captured in a  single Klein-Gordon field 
$X(z,t) = (1/\sqrt{L}) \sum_k e^{ i k z} \sqrt{\hbar/k} \left(
b_+(k+k_F, t) + b_-(k-k_F, t) \right) $, such
that $\partial^\mu \partial_\mu X = ( \partial_t^2/c^2 - \partial_z^2)X=0$. 
This enables us to rewrite the linearized action as 
\begin{equation}
S =   \int d^2 x \Bigl\{ \partial^\mu X^* \partial_\mu X  + i \hbar \bar \psi
\gamma^\mu \partial_\mu \psi \Bigr\},
\end{equation} 
which is the action for the transverse modes of a free relativistic $N=1$ superstring in $3+1$ dimensions
\cite{GSW}. In modern language, the Lorentz invariance 
of this action appears here as an emergent phenomenon at long wavelenghts, because the 
underlying theory is not Lorentz invariant. This is very similar with the way in which Lorentz invariance appears in 
string-bit models \cite{Bergman95}.
A second property of this action is, that the fermionic part has classically chiral symmetry, but quantum-mechaniclly suffers from a chiral anomaly. Whereas in string theory this is an unwanted feature, in our case it has a physical origin, because it comes about from the fact that the underlying microscopic theory does not conserve the chiral current $i \bar \psi \gamma^\mu \gamma^5 \psi$, and only conserves the current $\bar \psi \gamma^\mu \psi$ associated with the conservation of the total number of fermions.

\subsection{Nonlocal interaction}
The presence of a kelvon implies that neighbouring vortex cores are slightly
shifted with
respect to each other. This
effect decreases the fermionic hopping amplitude and results in a interaction
term that couples fermions on neighbouring sites of the form
\begin{equation}
\mathcal{H}' =  t \sum_{k k'} A(k')  \cos (k) c_k^\dagger c_k b_{k'}^\dagger
b_{k'}   .
\end{equation}
 Since this term breaks supersymmetry, we want to investigate the system
parameters for which it can be neglected, i.e., for which $A(k')\ll 1$.
To do so we consider a kelvon with a certain wavenumber $k$. The relative
distance between neighboring cores kan then be estimated to be
\begin{equation}
\delta r = \frac{R_{\rm TF}}{\sqrt{N_B}} \frac{k \lambda}{2}.
\end{equation}
From Eq. (\ref{vortexcore}) we know that for small distances the vortex core
can be modeled as a harmonic potential with width $\xi$. Hence, the fermionic
wavefunctions are gaussians with the same width. So we have to compute
\begin{eqnarray}
A(k) &=& 1- \frac{ \int d^2 {\bf r}  e^{ - \frac{{\bf r}^2}{2 \xi^2}} e^{ -
\frac{({\bf r} - \delta r_0)^2}{2\xi^2}} }{ \int d^2 {\bf r} e^{- \frac{{\bf r}^2
}{ \xi^2}}  }   
\\ 
&=&  1 -  e^{ -\frac{\delta r_0^2}{4 \xi^2} } \nonumber  
\simeq  \frac{\delta r_0^2}{4 \xi^2} \nonumber 
= \frac{ k^2 \lambda^2 R_{\rm TF}^4}{ 16 N_b \ell^4}, 
\end{eqnarray}
where we used the relation from Eq. (\ref{TFR}). From this same relation we see
that $(R_{\rm TF}/\ell)^4$ scales with the number of bosonic atoms $N_B$, such that
$A(k)$ is independent of $N_B$. We can estimate $R_{\rm TF}^4/N_B \ell^4$ to be of order
unity, such that the requirement for $A(k)$ to be small only depends on the
wavenumber $k$. If we identify this wavenumber with the Fermi momentum, i.e.,
\begin{equation}
\frac{k_F^2 \lambda^2}{16} \ll 1, 
\end{equation}
we obtain a restriction on the fermionic filling fraction which can be estimated
to
be
\begin{equation}
N_F < 0.1.
\end{equation}
From Fig. \ref{tuning2} we see that for most of the parameter space this
condition is fullfilled.

\section{Experimental signal}
It is an important question how the supersymmetry can be observed. Therefore we
need to distinguish between the question whether the hamiltonian is tuned to be
supersymmetric and whether the quantum ground state is supersymmetric, since it
is
possible that the ground state
can spontaneously break supersymmetry. We are primarely interested in the
situation that both the hamiltonian and the quantum ground state are
supersymmetric.

\subsection{Density measurements}
The two observables that are most easy to measure experimentally are the average
number of fermions at a site $N_F$ and the average number of kelvons $N_K$.  The
average fermion
number can be determined by usual absorpsion measurements. The number of kelvons
can be obtained from the
mean-square displacement $\langle r^2 \rangle = (1/N_s) \sum_i \langle x_i^2 + y_i^2 \rangle$ of the pancake vortices, which
can be measured
by imaging along the
$z$ direction the size of the circle within which the
vortex positions are concentrated \cite{Martikainen04}. Because
\begin{equation}
b_i^\dagger b_i = N_B \frac{x_i^2 + y_i^2}{R_{\rm TF}^2} - \frac{1}{2}, 
\end{equation}
this can directly be translated to the number of kelvons at a site. 
It is clear that in order to have a supersymmetric state, the kelvon and fermion
modes should have the same average occupation number, i.e., 
%\begin{equation}
%\langle b_k^\dagger
%b_k
%\rangle = \langle c_k^\dagger c_k \rangle,
%\end{equation}
%which directly implies
\begin{equation}
\langle b_i^\dagger
b_i
\rangle = \langle c_i^\dagger c_i \rangle.
\end{equation}
This allows us to devise an experimental measure for the proximity to the
supersymmetric point, which can be directly measured, namely
\[
\left(
 N_B \frac{ \langle r^2 \rangle}{R_{\rm TF}^2} -   N_{F} - {1}/{2}
\right)^2.    
\]
This quantity has an absolute minimum of zero at the
supersymmetric point, so that it's magnitude is a measure of the
deviation from supersymmetry.
We can extend this to higher order correlation functions. The condition that
\begin{equation}
\langle (b_i^\dagger b_i)^2 \rangle = \langle (c_i^\dagger c_i)^2 \rangle =
\langle c_i^\dagger c_i \rangle,
\end{equation}
can be used to prove that in order to have supersymmetry also the condition 
\begin{equation}
N_B^2 \frac{\langle r^4 \rangle}{R_{\rm TF}^4} = 2 N_F + \frac{1}{4} 
\end{equation}
should hold.
The quantiy $\langle r^4 \rangle$ can again be measured from the distribution of
the
measured vortex postions.

\begin{figure}[t]
\begin{center}
\includegraphics[scale=.3]{bubble.eps}
\includegraphics[scale=.3]{bubble_wigly.eps}
\end{center}
\caption{Diagrams that contribute to the dissipation of the vortex line. 
The solid lines represent bosonic propagators,
the dashed lines represent fermionic propagators. The left diagram is called
$\Pi_{KK} (k, \omega)$, 
the left diagram is called $\Pi_{KF}(k, \omega)$}.
\label{diagramsl}
\end{figure}

\subsection{Dissipation}
Another consequence of supersymmetry that can be directly measured is the
reduced dissipation. Dissipation in this context results in the vortex spiraling
out of the gas. \cite{Fedichev99, Duine04}.
The dominant part of the dissipation is given by the coupling to the kelvon
modes and the fermionic modes. The lowest order diagrams are given in Fig.\
\ref{diagramsl} and denoted by  $\Pi_{KK} (k, \omega)$ for the coupling to the
kelvon modes and 
 $\Pi_{KF}(k, \omega)$ when there is also coupling to the fermionic modes. The
imaginary part of these diagrams measures the dissipation. In order to be able
to know the dissipation away from the supersymmetric point we perform the calculation
for unequal dispersions $\epsilon_{K,F}(k)$ and unequal coupling constant
$U_{KK}$ and $U_{KF}$. We introduce the usual notation for the Bose-Einstein and Fermi-Dirac distribution functions
\[
N_B \lbrack\epsilon(k)\rbrack = \frac{1}{e^{\epsilon(k)/k_B T} - 1}, \quad N_F
\lbrack \epsilon(k) \rbrack = \frac{1}{e^{\epsilon(k)/k_B T} + 1}.
\]
The diagrams are then given by
\begin{widetext}
\begin{eqnarray}
\Pi_{KK} (k, i \omega) &=&  2 U_{KK}^2 \int \frac{d p}{2 \pi} \int \frac{d p'}{ 2
\pi} 
%\\ &&  
\frac{\Bigl(1+ N_B \lbrack \epsilon_K (p) \rbrack + N_B  \lbrack \epsilon_K(p')
\rbrack \Bigr) N_B \lbrack \epsilon_K(p+p' - k) \rbrack - N_B \lbrack \epsilon_K
(p) \rbrack N_B \lbrack \epsilon_K (p') \rbrack}
{ i \hbar \omega -  \epsilon_K(p) - \epsilon_K(p') + \epsilon_K(p+p' - k)} \nonumber \\ 
\Pi_{KF} (k, i \omega) &=& -  U_{KF}^2 \int \frac{d p}{2 \pi} \int \frac{d p'}{
2 \pi} 
%\\ && 
\frac{\Bigl(-1 - N_B \lbrack \epsilon_K (p) \rbrack + N_F  \lbrack
\epsilon_F(p') \rbrack \Bigr) N_F \lbrack \epsilon_F(p+p' - k) \rbrack + N_B
\lbrack \epsilon_K (p) \rbrack N_F \lbrack \epsilon_F (p') \rbrack }
{ i \hbar \omega -  \epsilon_K(p) - \epsilon_F(p') + \epsilon_F(p+p' - k)}, \nonumber 
\end{eqnarray}
where the minus sign in front of the expression for $\Pi_{KF} (k, i \omega)$
comes from the presence of the fermion loop. Note that due to the different combinatorial factors the diagram $\Pi_{BB}$ comes with an extra factor of two, which is lacking in the case of the diagram $\Pi_{BF}$. As a result, the two diagrams do not cancel exactly at the supersymmetric point, as we claimed previously \cite{Snoek05}. Instead, the dissipation is reduces by a factor 2 \cite{Bernard}.
The imaginary part of the diagrams gives the following expressions
\begin{eqnarray}
{\rm Im} \lbrack \Pi_{KK} (k,  \omega) \rbrack &=&  2 U_{KK}^2 \int \frac{d p}{2
\pi} \int \frac{d p'}{ 2 \pi} \Bigl\{
\delta( \hbar \omega  -  \epsilon_K(p) - \epsilon_K(p') + \epsilon_K(p+p' - k))
\times 
\nonumber \\ && \hspace{1cm} 
\left(\Bigl(1+ N_B \lbrack \epsilon_K (p) \rbrack + N_B  \lbrack \epsilon_K(p')
\rbrack \Bigr) N_B \lbrack \epsilon_K(p+p' - k) \rbrack - N_B \lbrack \epsilon_K
(p) \rbrack N_B \lbrack \epsilon_K (p') \rbrack \right) \Bigr\}, \nonumber \\ 
 {\rm Im} \lbrack \Pi_{KF} (k,  \omega) \rbrack &=& -  U_{KF}^2 \int \frac{d p}{2
\pi} \int \frac{d p'}{ 2 \pi} \Bigl\{
\delta(\hbar  \omega  -  \epsilon_K(p) - \epsilon_F(p') + \epsilon_F(p+p' - k))
\times 
\nonumber \\ && \hspace{1cm}
\left(\Bigl(-1 - N_B \lbrack \epsilon_K (p) \rbrack + N_F  \lbrack
\epsilon_F(p') \rbrack \Bigr) N_F \lbrack \epsilon_F(p+p' - k) \rbrack + N_B
\lbrack \epsilon_K (p) \rbrack N_F \lbrack \epsilon_F (p') \rbrack \right).
\end{eqnarray}
\end{widetext}
At zero temperature we have that $N_F \lbrack \epsilon(k) \rbrack = \theta(\epsilon(k))$,
and $N_B \lbrack \epsilon(k) \rbrack =  - \theta(\epsilon(k)) = - N_F \lbrack
\epsilon(k) \rbrack$. Using this, we see that if there is supersymmetry, i.e.,
if $\epsilon_K (k) = \epsilon_F (k)$ and $U_{KK} = U_{KF}$, we have that 
\[\Pi_{KK} (k, \omega)  = - 2 \Pi_{KF} (k, \omega), \]
and in particular that 
\[{\rm Im} \lbrack \Pi_{KK} (k, \omega) + \Pi_{KF} (k, \omega) \rbrack = \frac{1}{2} {\rm Im} \lbrack \Pi_{KK} (k, \omega) \rbrack,\]
such that at zero temperature supersymmetry results in a dissipation rate that is only half as large as in the case of a ordinary vortex-line.
Using these expressions, it is also possible to calculate the quantum
dissipation at nonzero temperature, or when supersymmetry is broken.
In particular, when the interaction coefficients are tuned away from the supersymmetric point such that $U_{BF} = \sqrt{2} U_{BB}$ and supersymmetry is maintained at the quadratic level, the dissipation exactly vanishes and the superstring is extremely stable in the center of the condensate.

\subsection{Spontaneous supersymmetry breaking}
When the hamiltonian is supersymmetric, the ground state still can break supersymmetry. This is the phenomenon of spontaneous supersymmetry breaking. For $\Omega< \Omega_c$, the ultracold superstring is unstable against
Bose-Einstein condensation of kelvons. This breaks supersymmetry, because the fermionic modes cannot Bose-Einstein condense.
Bose-Einstein condensation implies that the kelvon annihilation operator obtains an expectation value
\begin{equation}
 \langle b_i \rangle  \neq 0. 
\end{equation}
From the definition of the kelvon operator we conclude that as a consequence 
\begin{equation}
\langle x \rangle^2 + \langle y \rangle^2 > 0. 
\end{equation}
This means that the vortex moves out of the center of the trap. Experimentally this is easy to measure.
Moreover, by monitoring the vortex position when it moves out of the center of the trap, this also allows for the 
experimental investigation of the dynamics of supersymmetry breaking.
As a consequence of the breaking of the $U(1)$ symmetry because of the Bose-Einstein condensation, the dispersion of the kelvon modes becomes gapless. The dispersion becomes the usual Bogoliubov dispersion, which reads
\begin{equation} \label{bosdisp}
\hbar \omega_B(k) = \sqrt{\epsilon(k)^2 + 2 \mu \epsilon(k)},
\end{equation}
with $\epsilon(k) = \hbar^2 k^2/2m^*$. For long wavelengths this yields a linear behaviour. Also the fermionic modes become gapless. This is a result of the breaking of supersymmetry and this mode is called the goldstino. Because $\langle b \rangle = \sqrt{\mu/U}$, the dispersion for the goldstino is given by
\begin{equation} \label{fermdisp}
\hbar \omega_F(k) = \epsilon(k)  - \mu + |\langle b \rangle|^2 = \epsilon(k),
\end{equation}
which results in a quadratic dispersion. Clearly the bosonic and fermionic dispersion in Eqs. \eqref{bosdisp} and \eqref{fermdisp} are now different, which signals a nonsupersymmetric situation. 

% \subsection{Optimization of fermionic density}
% The typical fermionic number of particles is around $0.1$ per site. This means
% for
% a total of 1000 sites that there are around 100 fermions in the systems. This is
% very little, both to control and to observe. 
% However, the density is rather high and can be estimated to be at least
% $10^{13}/{\rm cm}^2$. Moreover, a disadvantage of a higher fermionic atomic
% density is that this makes also the contribution of the kelvon-fermion hopping
% interaction more important. 
%However, a compromise is very well possible.
%Ways to crank op the fermion number are:
%1) Tuning close to resonance from below: then the fermionic density goed up to
%one.
%2) Play with the bosonic number of particles: bringing it down leads to weaker
%interactions, such that the Thomas-Fermi anstatz fails more and more. We should
%do the calculation to see how this improves the fermion number.
%3) Play with the rotation: rotating the other way around certainly helps, but
%makes the sytem also more unstable. However, supersymmetry protects the system
%against spiraling out!

\section{Supersymmetry}
In this section we review the algebra associated with supersymmetric field
theories both in the relativistic (super Poincar\'e algebra) and the
non-relativistic limit (super Galilei algebra). We give an explicit
representation of the super Galilei algebra in terms of the bosonic and
fermionic operators.

\subsection{Super Poincar\'e algebra}
Associated with a relatistic field theory in $D=d+1$ dimensions is the Poincar\'e
algebra, whose generators consist of a vector $P^\mu$, that generates
translations and an antisymmetric tensor $J^{\mu \nu}$, that generates Lorenz
transormations. The greek indices run from $0$ to $d = D-1$, such that
$P^0$ should be identified with the hamiltonian $\mathcal{H}$, up to a
constant. 
The algebra is then given by
\begin{eqnarray}
\lbrack P^\mu, P^\nu \rbrack &=& 0 \\
\lbrack J^{\mu \nu}, P^{\rho}\rbrack &=& i (\eta^{\mu \rho} P^{\nu} -  \eta^{\nu
\rho} P^{\mu}) \nonumber \\
\lbrack J^{\mu \nu},  J^{\rho \sigma} \rbrack &=& i (\eta^{\mu \rho} J^{\nu
\sigma} + \eta^{\mu \sigma} J^{\rho \nu }
- \eta^{\nu \rho} J^{\mu \sigma} - \eta^{\nu \sigma} J^{\rho \mu}) \nonumber,
\end{eqnarray}
where $\eta^{\mu \nu}$ is the flat space Minkowski metric.
When there is supersymmetry we can extend this to the super Poincar\'e algebra.
For $N=1$ supersymmetry in $1+1$ dimensions this involves the two-component Majorana spinor $Q_\alpha$, $\alpha=1,2$, which is
the generator of
supersymmetry transformations. The algebra is then extended to include also
\begin{eqnarray}
\lbrack P^\mu, Q_\alpha \rbrack &=& 0\\
\lbrack P^\mu, \bar Q_\alpha \rbrack &=& 0\\
\lbrack J^{\mu \nu}, Q_\alpha \rbrack &=& - \frac{i}{4} \lbrack \gamma^{\mu},
\gamma^{\nu} \rbrack_{\alpha \beta} Q_\beta \\
\{ Q_\alpha, \bar Q_\beta \} &=& 2 \gamma_{\alpha \beta}^\mu P_\mu,
\end{eqnarray}
where the $\gamma^{\mu}$ are again the Dirac matrices.
We use conventions such that $Q_\alpha$ has two real components. To make connection with the supersymmetry in the ultracold superstring
we combine these two components in one complex supersymmetry operator
\begin{equation}
Q = \frac{Q_1 +  i Q_2}{2}.
\end{equation}
This decomposition breaks manifest Lorentz symmetry, but since we are ultimately interested in the nonrelativistic limit, this is of no concern to us here. As a result we obtain the following algebra  
\begin{eqnarray}
\lbrack P^\mu, Q \rbrack &=& 0\\
\lbrack J^{\mu \nu}, Q  \rbrack &=& -\frac{i}{2} \epsilon^{\mu\nu} Q^\dagger \\
\{ Q, Q^\dagger \} &=&  P^0 \\
\{ Q, Q \} &=& \{ Q^\dagger, Q^\dagger \}=- P^1.
\end{eqnarray}
In particular, we see that the hamiltonian $P^0$ is fixed by the supersymmetry
generator. This is a very peculiar restriction on the hamiltonian, which is only
true for the relativistic theory. In the nonrelativistic limit, the supersymmetry
decouples from the space-time translation symmetry as we show now.

\subsection{Super Galilei algebra}
The Galilei algebra can be derived as a limit of the Poincar\'e algebra by
performing a 
In\"on\"u-Wigner contraction \cite{Inonu53} in the
following way \cite{Puzalowski78, Clark84}. We write
\begin{eqnarray}
P^0 &=& \frac{1}{c} ( m^* \mathcal{N} c^2 + \mathcal{H}) \\
P^1 &=& P \\
J^{0 1} &=&  c K \\
Q &=& \sqrt{c} Q,
\end{eqnarray}
where $c$ is the speed of light and $m^*$ denotes the mass, which is the same for the bosonic and
fermionic degrees of freedom.  
We also defined a number
operator $\mathcal{N}$, which counts all the particles in the system, and boost operators
$K$. Furthermore, we still have the space
translation generators $P$  and the
hamiltonian $\mathcal H$. We can now take the limit $c \rightarrow \infty$ to
obtain the super Galilei algebra.
The Galilei algebra obtained in this manner has nonvanishing commutators
\begin{eqnarray}
\lbrack P, K \rbrack &=& i  m^* \mathcal{N} \\
\lbrack \mathcal{H}, K \rbrack &=& i P. 
\end{eqnarray}
The part involving the supersymmetry becomes only
\begin{equation}
\{ Q, Q^\dagger \} = \mathcal{N}.
\end{equation} 
This defines the algebra $\mathcal{S}_1\mathcal{G}$ \cite{Bergman95}. As is 
clear, in this case the hamiltonian is decoupled form the supersymmetry.
In $1+1$ and $2+1$ dimensions, it is sometimes possible to define an extended
superalgebra $\mathcal{S}_2\mathcal{G}$, which again involves the hamiltonian
\cite{Bergman95, Bergman95B, Bergman96}. In $d=1$ this amounts to introducing an extra
scalar supersymmetry generator $R$ with the algebra:
\begin{eqnarray}
\{Q, R^\dagger \} &=& - P\\
\{ R, R^\dagger\} &=& \mathcal{H}/2.
\end{eqnarray}

\subsection{Representation}
The representation for the $\mathcal{S}_1 \mathcal{G}$ algebra in terms of the
bosonic and fermionic operators $b$ and $c$, can easily be found to be
\begin{eqnarray}
\mathcal{N} &=& \int d z \{ b^\dagger (z) b(z) + c^\dagger (z) c(z) \}, \\
P &=& - \frac{i\hbar}{m^*} \int d z \{  b^\dagger (z) \partial_z b(z) +
c^\dagger
(z) \partial_z c(z) \}, \\ 
Q &=&  \int d z \;  c^\dagger (z) b(z) , \\
Q^\dagger &=&  \int d z \;  b^\dagger (z) c(z). \\
\end{eqnarray}
In addition, we can thus also define 
\begin{eqnarray}
R &=& \frac{\hbar}{\sqrt{m^*}} \int dz \; c^\dagger (z) \partial_z b(z),  \\
R^\dagger &=&  -  \frac{\hbar}{\sqrt{m^*}} \int dz \; b^\dagger (z) \partial_z
b(z).
\end{eqnarray}
This produces 
\begin{equation}
\{ R, R^\dagger \} = \int d z \left\{- b^\dagger (z) \frac{\hbar^2}{ m^*}
\frac{\partial^2}{\partial z^2} b (z) - 
c^\dagger (z)   \frac{\hbar^2}{m^*}  \frac{\partial^2}{\partial z^2}
 c(z) \right\},
\end{equation}
which indeed is the kinetic energy part of the hamiltonian. The full quadratic part of the hamiltonian
can be expressed as
\begin{equation}
\mathcal{H}= \frac{1}{2} \{ R, R^\dagger \} - \mu \{ Q, Q^\dagger \}.
\end{equation}
%It is possible to extend this analysis to include also the interaction terms. 

For completeness, we mention that we can also use superspace techniques to write the hamiltonian in
a manifest supersymmetric way. This involves the introduction of a complex superfield
\begin{eqnarray}
\Psi(z, \theta) &=& e^{- \theta^* \theta/2}(b(z) + \theta^* c(z))\\
\Psi^* (z, \theta) &=& e^{-\theta^* \theta^*/2}(b^*(z)  + c^* (z) \theta), 
\end{eqnarray}
where $\theta$ is a Grassman variable such that $\{\theta, \theta\}= \{\theta^*,
\theta^*\}=0$ and $\{\theta, \theta^*\}=0$. The hamiltonian is in terms of the
superfield given by
\begin{eqnarray}
\mathcal{H} &=& \int d \theta^* d \theta \int d z \Bigl\{  \frac{\hbar^2}{2m^*}
|\partial_z \Psi (z, \theta)|^2 - \mu |\Psi (z, \theta)|^2 \nonumber \\  
&& \hspace{3cm} + \frac{U}{2} |\Psi (z, \theta)|^4 \Bigr\}.
\end{eqnarray}
In this formulation the spontaneous breaking of supersymmetry is particularly elegant, because the hamiltonian has the form of a standard Landau theory of a second-order phase transition with $\langle \Psi(z, \theta)\rangle$ as the order parameter.

% We also introduce the superspace operators 
% \begin{eqnarray}
% D_\pm &=& \mp \frac{\partial}{\partial \theta} + \frac{1}{2} \theta^*, \\
% \bar D_\pm &=& \frac{\partial}{\partial \theta^*} \pm \frac{1}{2} \theta,
% \end{eqnarray}
% which obey the following identities:
% \begin{eqnarray}
% (D_\pm)^2 &=& (\bar D_\pm)^2 = 0, \\ 
% \{D_+, D_-\} &=& \{D_+, \bar D_+\} = \{D_-, \bar D_-\} = 0 \\
% \{D_+, \bar D_-\} &=& \{D_- , \bar D_+\}=1.
% \end{eqnarray}
% The supersymmetry transformation on the original fiels are:
% \begin{equation}
% \delta b = \epsilon^* c, \quad \delta c = \epsilon b.
% \end{equation}
% This translates into the following transformation of the superfield:
% \begin{equation}
% \delta \Psi = (\epsilon D_- + \epsilon^* \bar D_+)\Psi.
% \end{equation}
% Using the identities it is then straight-forward to derive that 
% \begin{equation}
% \delta {H} =0.
% \end{equation}

\section{Connection with string theory}
In this section, we discuss the similarities and differences with
superstring theory. For some textbooks on the subject, we refer to Refs.
\cite{GSW, LT, P}. In string theory, one usually starts with the Polyakov
action \cite{Pol}, describing the coordinates
$X^{\mu}(\sigma,\tau)$, with $\, \mu =0,1,...,D-1$, of the string
propagating in a $D$-dimensional curved space-time with metric
$G_{\mu\nu}(X)$,
\begin{equation}\label{action}
S=-\frac{T}{2}\int\,{\rm d}^2\sigma\,{\sqrt h}h^{\alpha\beta}\partial_\alpha
X^\mu \partial_\beta X^\nu G_{\mu\nu}(X)\ .
\end{equation}
Here $\sigma^\alpha=\{\sigma, \tau \}$ are coordinates on the
worldsheet sweeped out by the string, $\tau$ is the worldsheet
time, and $\sigma$ runs longitudinally over the string.
Furthermore, $T$ is the string tension, and $h_{\alpha\beta}$ is a
two-dimensional metric on the worldsheet with $h=-{\rm Det}\lbrack
h_{\alpha\beta} \rbrack$. In agreement wit the standard practice in high-energy physics we are momentarily using units such that $\hbar = c = 1$. We restore units when we come to the precise connection with our ultracold superstring.
In fully quantized string theory, one also performs a
path integral over these metrics, and this leads to the string
loop expansion where one sums over all two-dimensional surfaces
containing an arbitrary number of holes. In our setup, the
worldsheet of the string is completely fixed, and contains no
holes, i.e., it is just the two-dimensional plane. On the plane,
we can then make use of the local symmetries  of the Polyakov action, that are the reparameterizations
of the worldsheet coordinates and the Weyl rescalings of the metric. Doing so, we can make the gauge choice
\begin{equation}\label{conf-gauge}
h_{\alpha\beta}=\left(\begin{matrix} 1 & 0 \cr 0 & -1\end{matrix}\right)\ .
\end{equation}
This gauge choice is referred to as the conformal gauge.

The space-time in which the string propagates is coordinatized by
$X^\mu,\, \mu=0,...,D-1$. In quantized superstring theory one has
that $D=10$, but at the classical level one can have $D=4$ as
well. We come back to this issue below. It is useful to introduce
light-cone variables
\begin{equation}
X^\pm=\frac{1}{\sqrt 2}(X^0\pm X^{D-1}) \; ,
\end{equation}
and $
X^i, \quad i=1, \ldots ,D-2$.
Then $X^\pm$ and $X^i$ describe the longitudinal and transversal
degrees of freedom of the string, respectively. String theory has the special feature
that there are only transversal physical degrees of freedom. This
is because string theory has an additional constraint that can be
understood as the equation of motion of the worldsheet metric
$h_{\alpha\beta}$. Defining $\sigma^\pm=\tau \pm \sigma$, these
constraints read in conformal gauge
\begin{equation}\label{constr}
\partial_\pm X^\mu \partial_\pm X^\nu G_{\mu\nu}(X)=0\ ,
\end{equation}
and are sometimes called the Virasoro constraints.
In practice, solving the constraints is difficult, but in the so-called
light-cone gauge
\begin{equation}\label{light-cone}
X^+=2\alpha' p^+ \tau, 
\end{equation}
where $\alpha'\equiv (2\pi T)^{-1}$ and $p^+$ is the center-of-mass momentum in the $X^+$ direction,
the longitudinal modes $X^\pm$ can be eliminated explicitly, at
least for certain space-time metrics $G_{\mu\nu}$. The light-cone
gauge can always be taken as a consequence of the residual gauge symmetry
after the gauge choice of Eq.\ \eqref{conf-gauge} has been imposed \cite{GSW, LT, P}.
 %For more details, see e.g. \cite{GSW}.

The implementation of the constraints in Eq.\ \eqref{constr} in the quantum theory
leads to the critical dimension, namely $D=26$ for the bosonic string and
$D=10$ for the superstring. In our condensed-matter setup, these constraints
are not present. There are physical longitudinal degrees of freedom, so this
makes it different from the superstring. However, the
longitudinal modes are suppressed and at the energy
scales we are looking at, it suffices to study only the transversal degrees.
It is in this transversal sector that we  connect to string theory.
To make this connection, we have to specify the space-time
metric $G_{\mu\nu}$. A class of backgrounds that has been intensely studied
in the string literature is that of plane wave metrics \cite{PW, RT}.
% see e.g. \cite{RT}, or \cite{PW} for the original
% references. 
The simplest of these backgrounds, and also the one
relevant for our case, is given by
\begin{eqnarray}\label{wave}
{\rm d}s^2 &\equiv& G_{\mu\nu}(X){\rm d}X^\mu{\rm d}X^\nu \\ &=&
-2{\rm d}X^+{\rm d}X^- + H(X^i) ({\rm d}X^+)^2 + {\rm d}X^i{\rm d}X^i, \nonumber
\end{eqnarray}
where $H(X^i)$ is a function of the transverse coordinates only.
In light-cone gauge, the lagrangian for the string propagating in this
background now becomes
\begin{equation}\label{lagr-pot}
T^{-1}{\cal L}= \frac{1}{2} \sum_{i=1}^{D-2} \left[ \left(\frac{\partial X^i}
{\partial \tau}\right)^2-\left(\frac{\partial X^i}{\partial
\sigma}\right)^2\right] - V(X^i)\ ,
\end{equation}
where $V(X)=-2(\alpha'p^+)^2H(X)$. To derive this result, we simply
substitute the background  in Eq.\ \eqref{wave} into Eq.\ \eqref{action}, and use the
light-cone gauge from Eq.\ \eqref{light-cone} to produce the potential $V(X)$ term in
the lagrangian~\footnote{That the light-cone gauge can be imposed
simultaneously with the conformal gauge from Eq.\ \eqref{conf-gauge} was
shown in Ref. \cite{HS}.
In fact, the only space-times in which this can be done are flat space
and the plane-wave backgrounds.}. Furthermore, this produces a term
proportional to $X^-$ that
is decoupled from the $X^i$. Therefore this term can be dropped. In fact $X^-$
is fixed by the Virasoro constraints in Eq.\ \eqref{constr}, so we only need
a lagrangian for the transverse degrees of freedom.

One of the remarkable facts of string theory is that its conformal
symmetry at the quantum level forces the metric to satisfy Einstein's
equations in general relativity. This is
the way in which gravity emerges from string theory. When there are
no other background fields present, as in our case, Einstein's equations
reduce to a single constraint on the function $H$ given by
\begin{equation}\label{potential}
\Delta H \equiv \sum_{i=1}^{D-2}\frac{\partial^2 }{(\partial X^i)^2}
H=0\ .
\end{equation}
In other words, $H$ has to satisfy the Laplace equation in the
transverse space. This constraint has to be understood on an equal
footing as the constraint on the space-time dimension. They both
follow from a consistent implementation of the conformal symmetry
at the quantum level. Since we are not taking into account the
Virasoro constraints in our system, and hence the conformal
symmetry, we therefore also ignore the constraint in Eq.\
\eqref{potential}.
% , in the same way as we ignored the critical
% dimension. 
Doing so, we can work with arbitrary potentials $V(X)$.
When we take $D=4$, as we shall below, the scalar potential
depends on two real fields.

We now include the fermions, and discuss supersymmetry. To make a
superstring we have to add additional terms to the lagrangian in Eq.\
\eqref{action} containing the fermions in such a way that there is
supersymmetry. We can then impose the conformal or light-cone gauges
to arrive at a supersymmetric generalization of the lagrangian in Eq.\
\eqref{lagr-pot}. Alternatively, we can directly study
supersymmetric extensions of Eq.\ \eqref{lagr-pot} as two-dimensional
field theories. The general construction of supersymmetric
two-dimensional field theories with scalar potentials $V(X)$ was
given in Ref. \cite{AGF}. Not all potentials lead to lagrangians that
can be supersymmetrized. For the case of minimal supersymmetry
with two supercharges, sometimes denoted by (1,1) SUSY, the potential
needs to be of the following type
\begin{equation}\label{V-pot}
V(X^i)=\sum_{i=1}^{D-2} \left(({\partial_i W})^2
+G^2_i(X)\right)\ ,
\end{equation}
where $W$ is a real function, $\partial_i$ stands for the
derivative with respect to $X^i$, and the quantities $G_i(X)$
satisfy $\partial_iG_j+\partial_jG_i=0$ together with $\sum_i
G^i\partial_i W = {\rm constant}$. The supersymmetric lagrangian
can then be written as
\begin{eqnarray}\label{susy-lag}
2 T^{-1} {\cal L} &=& \partial_\alpha X^i \partial^\alpha X^i + i{\bar
\psi}^i\gamma^\alpha \partial_\alpha \psi^i - V(X) \\ && \nonumber -W_{ij}(X)
{\bar \psi}^i \psi^j - W^{(5)}_{ij}(X)\,{\bar \psi}^i\gamma^5
\psi^j,
\end{eqnarray}
with 
\begin{equation}
W_{ij}=\partial_i\partial_j W\ ,\qquad W^{(5)}_{ij}=\partial_i
G_j.
\end{equation}
The supersymmetry variations are
\begin{eqnarray}
\delta X^i &=& {\bar \epsilon}\psi^i, \\ 
\delta \psi^i &= &-i\gamma^\alpha\partial_\alpha X^i\epsilon
-\partial^iW \epsilon - G^i \gamma^5 \epsilon,
\end{eqnarray}
and leave the lagrangian invariant, up to a total derivative. Here
$\psi^i$ and $\epsilon$ are two-component Majorana spinors, and in
our model we thus have two Majorana spinors. The $\gamma$-matrices are
related to the Pauli matrices as $\gamma^0=\sigma_y,
\gamma^1=i\sigma_x$ and $\gamma^5=\sigma_z$ as before. For more details on
the spinor conventions, see Ref. \cite{AGF}.

Examples of supersymmetric models are given by
\begin{equation}
G_i=\alpha \epsilon_{ij}X^j\ ,\qquad W(r)=\beta R +\gamma R^3\ ,
\end{equation}
where $R \equiv {\sqrt {(X^1)^2+(X^2)^2}}$ and
$\alpha,\beta,\gamma$ are arbitrary parameters. Plugging this into
Eq.\ \eqref{V-pot} leads to~\footnote{One could also add a term
proportional to $R^2$ to $W$. This leads, however, to terms in the potential
$V$ with odd powers in $R$, which is not what we are looking for.}
\begin{equation}\label{mh-pot}
V(R)=\left(\beta +3\gamma R^2\right)^2+\alpha^2 R^2.
\end{equation}
Up to an irrelevant additive constant, the coefficients $\alpha,\beta,\gamma$
can be chosen such that the potential is as in our
condensed-matter setup. Furthermore, we have that $W^{(5)}_{ij}=
\alpha\epsilon_{ji}$
which leads to mass terms for the fermions, and supersymmetry variations of the fermions
of the form
\begin{equation}
\delta \psi^i = \cdots + \alpha \epsilon^{ij}X_j \gamma^5 \epsilon\ .
\end{equation}
This term rotates the fermions into the bosons, just like for the ultracold
superstring. If we compute $W_{ij}$
to determine the interactions between bosons and fermions, it produces
complicated interaction terms,
\begin{equation}
W_{ij}=\delta_{ij} \left(\frac{\beta}{R}+3\gamma R \right) +X_iX_j
\left(-\frac{\beta}{R^3}+\frac{3\gamma}{R}\right)\ ,
\end{equation}
as a result of the supersymmetry constraints.

%Moreover, it leads to additional terms in the supersymmetry
%transformations of the fermions. These terms 
% could perhaps
%disappear after taking the nonrelativistic limit, which we
%discuss now.

\subsection{Nonrelativistic limit}
To connect to our condensed-matter setup, we have to take the
nonrelativistic limit in which only particle excitations of the
two-dimensional field theory survive, and the anti-particle
excitations are absent. To illustrate this procedure, we start
with the bosonic part of the lagrangian in Eq.\ \eqref{susy-lag}, based on
two real scalar fields. In terms of the complex field
\begin{equation}
X=X^1+iX^2\ ,\qquad  X^*=X^1-iX^2\ ,\qquad R^2 = |X|^2 ,
\end{equation}
the lagrangian reads
\begin{equation}\label{lagr-X}
T^{-1} {\cal L}=\frac{1}{c^2}|\partial_\tau X|^2 - |\partial _\sigma
X|^2- V(|X|)\ ,
\end{equation}
where we have reinserted the speed of light $c$ in order to take
the nonrelativistic limit $c\rightarrow \infty$ below, and we
further used that the potential is a function of $R$ only since
this is the case of interest. 

Using now the mass $m^*$, we decompose the complex scalar field in terms of positive and
negative frequency modes
\begin{equation}\label{decomp}
X(\sigma, \tau)=\frac{1}{\sqrt {2{m^* T}}}\left({\rm e}^{-i{m^*}c^2\tau}
b(\sigma,\tau)+{\rm e}^{i{m^*}c^2\tau }a(\sigma,\tau)\right),
\end{equation}
and call $b$ the particle field and $a$ the antiparticle field.
Both $b$ and $a$ are complex. We now substitute Eq.\ \eqref{decomp} into
Eq.\ \eqref{lagr-X} and send $c\rightarrow \infty$. In this limit, the
lagrangian becomes first order in time derivatives, and particles
and antiparticles decouple from each other such that we can
effectively set $a=0$. The remaining terms in the nonrelativistic
limit are
\begin{equation}
{\cal L}= i \hbar b^*\partial_\tau b -\frac{\hbar^2}{2{m^*}}|\partial _\sigma b|^2- V(|b|),
\end{equation}
where we have reinserted the various factors of $\hbar$.
Moreover, we have absorbed a mass term proportional to
$|b|^2$ into the potential. Remind that we have chosen a
potential of the form given in Eq.\ \eqref{mh-pot}, so this mass term
can easily be absorbed into a change of the coefficients $\alpha$
or $\beta\gamma$. Notice that this lagrangian precisely coincides
with the bosonic sector of lagrangian of the ultracold superstring given in Eq. \eqref{lagrangian}. 
The fermionic sector can be obtained in a similar way.
% 
% What remains to work
% out is the non-relativistic limit of the fermionic sector, but
% these terms should follow directly from imposing non-relativistic
% supersymmetry.

\section{Conclusion}
In this paper we presented a detailed account of the conditions under which the 
ultracold superstring can be created. The requirements for the laser parameters 
and the atomic interactions were given. Moreover we payed attention to the
experimental signatures of supersymmetry.
The supersymmetry in the problem was investigated by studying the appropriate super algebra.
Finally, a precise  mathematical connection with string theory in $3+1$ dimensions was made.

The discussions in this article were limited to the case of a single string. It
is left for future investigation to extend the analysis to involve more strings.
A complication in this case is that for parallel vortex lines, supersymmetry is
not possible, because of the different way vortices and fermions interact with
each other. A proposal to overcome this problem is to study the interaction of
two superstrings that are both in the center of the condensate, but are
seperated on the $z$ axis. This would correspond to merging and splitting of
ultracold superstrings.

The typical fermionic number of particles that is needed to obtain supersymmetry is typically around $0.1$ per site. This is
rather low, both to control and to observe. 
However, the density is rather high and can be estimated to be at least
$10^{13}\; {\rm cm}^{-3}$. Moreover, a disadvantage of a higher fermionic atomic
density is that this makes also the contribution of the kelvon-fermion hopping
interaction more important.  It remains to be investigated, whether a change of the varous parameters can improve on this situation.

Apart from the other possibilities mentioned in this article, it is also possible to gain experimental insight in the system by coupling the vortex motion to resonant quadrupole modes \cite{Martikainen04}. This gives the possibility to measure the kelvon dispersion directly. If the system is brought out of equilibrium by populating a high-lying kelvon mode, it also opens up the exciting possibility to study collapse and revival phenomena between the bosonic and fermionic modes. 

\section*{Acknowledgements}
We are grateful for helpful discussions with Masud Haque, Randy Hulet, Jan Ambj{\o}rn and Bernard de Wit.
This work is supported by the Stichting voor
Fundamenteel Onderzoek der Materie (FOM) and the Nederlandse
Organisatie voor Wetenschappelijk Onderzoek (NWO).

\end{document}